\documentclass{aastex}
\usepackage{natbib}
\usepackage{enumitem}
\usepackage{booktabs}
\usepackage{longtable}
\usepackage{multicol}
\usepackage{lscape}

\begin{document}
\title{Distant Solar System Objects identified in the Pan-STARRS1 survey}

\author{R.J. Weryk\altaffilmark{1}, E. Lilly\altaffilmark{1}\\
	S. Chastel\altaffilmark{1}, L. Denneau\altaffilmark{1},
	R. Jedicke\altaffilmark{1}, E. Magnier\altaffilmark{1}, R.J. Wainscoat\altaffilmark{1}\\
	K. Chambers\altaffilmark{1}, H. Flewelling\altaffilmark{1},
	M.E. Huber\altaffilmark{1}, C. Waters\altaffilmark{1}\\
	PS1 Builders}

\slugcomment{32 Pages, 9 Figures, 7 Tables}
\altaffiltext{1}{Institute for Astronomy, University of Hawaii, 2680 Woodlawn Drive, Honolulu HI 96822, USA}
\shorttitle{Distant Object search with Pan-STARRS1}
\shortauthors{Weryk\etal}

\begin{abstract}

We present a method to identify distant solar system objects in long-term wide-field asteroid
survey data, and conduct a search for them in the Pan-STARRS1 (PS1) image data acquired
from 2010 to mid-2015.  We demonstrate that our method is able to find multi-opposition
orbital links, and we present the resulting orbital distributions which consist of $154$
Centaurs, $255$ classical Trans-Neptunian Objects (TNOs), $121$ resonant TNOs, $89$
Scattered Disc Objects (SDOs) and $10$ comets.  Our results show more than half of these
are new discoveries, including a newly discovered 19th magnitude TNO.  Our identified
objects do not show clustering in their argument of perihelia, which if present, might
support the existence of a large unknown planetary-sized object in the outer solar system.

\end{abstract}
\maketitle


\section{Introduction}

\subsection{Background and Importance}

\begin{figure}[p]
\centering
\includegraphics[width=0.9\textwidth]{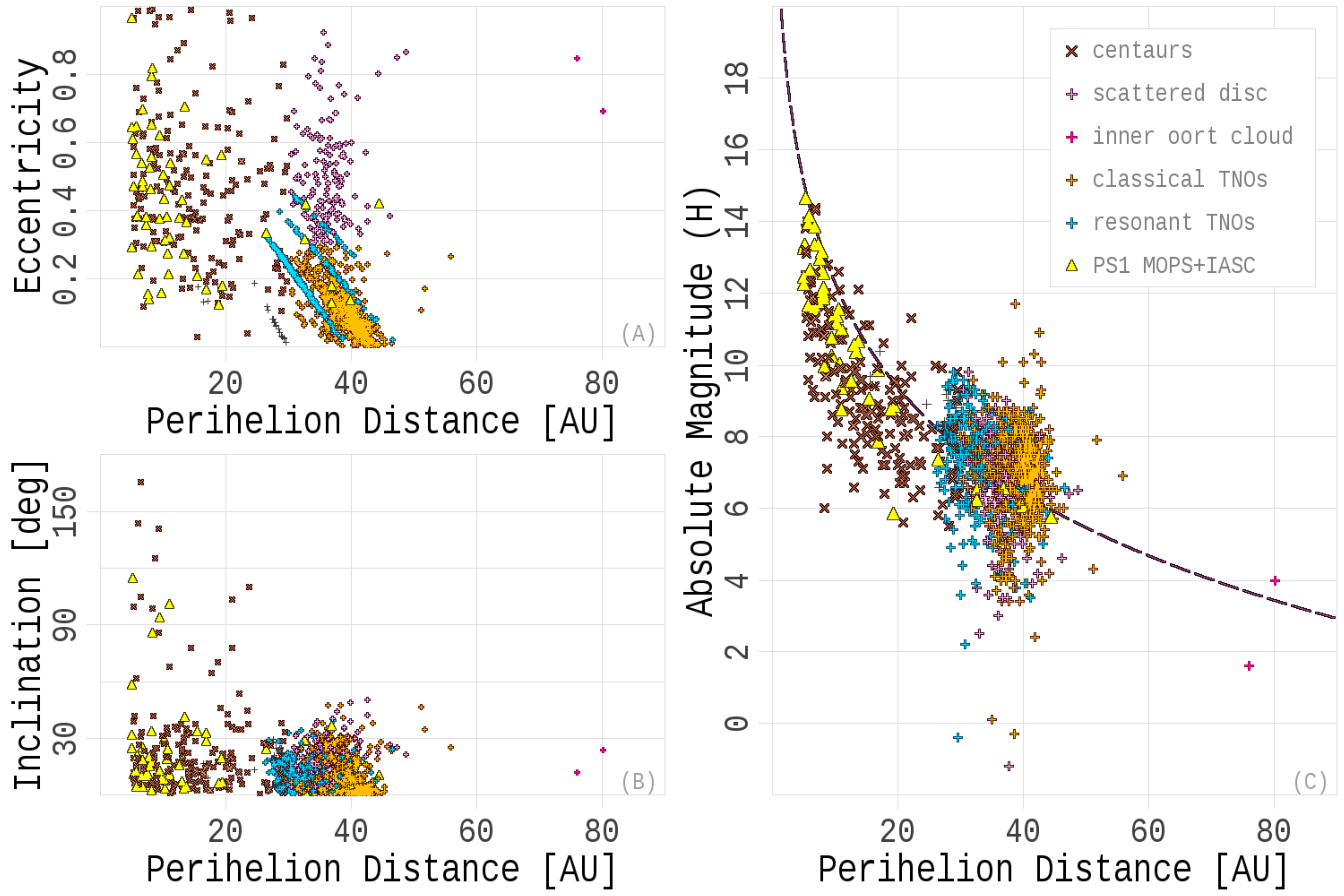}
\caption{(A) Eccentricity, (B) inclination, and (C) absolute magnitude, versus perihelion distance
of known TNOs (including SDOs; orange, magenta, and blue marks), Centaurs (brown), and Inner Oort Cloud objects
(IOCs; red dots). Objects previously discovered by PS1 through MOPS and IASC (as mentioned in the text)
are depicted as yellow triangles.  The limiting absolute magnitude is shown as a dashed line in panel
(C) for $V=22.5$.}
\label{real}
\end{figure}

The minor planets orbiting beyond Neptune provide valuable insight on our solar system's
formation and evolution, but they have only been studied since 1992 when the first
Trans-Neptunian Object (TNO) after Pluto was discovered \citep{Luu1993}.  Almost a quarter
century later, $\sim2000$ TNOs and Centaurs are known (see Figure \ref{real}) and they are
revealing their properties slowly because of the difficulties involved with detecting the
faint, slow-moving members of this distant population.

Multiple dedicated TNO surveys have been conducted over the years \citep[e.g.,][]{Larsen2001,Gladman2001,
Bernstein2004,Elliot2005,Petit2006,Petit2008,Sheppard2011,Gladman2012,Alexandersen2014,Brown2015}
including stellar occultation surveys \citep[e.g.,][]{Schlichting2012} which focussed on the discovery
of sub-km objects below the sensitivity limit of optical telescopes.  Thanks to these studies, the
large $100-1000$~km objects have a well characterized size-frequency distribution \citep[SFD;][]
{Petit2008,Fuentes2008} while TNOs smaller than $100$~km have only more recently been studied
\citep{Fraser2009,Sheppard2010b,Gladman2012}.  However, there is a need for more observational
data to confirm the apparent transition from a steep to shallow SFD slope among the Neptune Trojans
\citep{Sheppard2010b} and SDOs \citep{Shankman2013} around $D\sim100$~km (corresponding to absolute
magnitude $H\sim8.5$) \citep{Alexandersen2014}.  If the transition is present within all TNO
sub-populations it would suggest the formation scenario in which ``asteroids were born big''
\citep{Morbidelli2009} and imply that objects smaller than $100$~km are dominantly the result
of collisional evolution.

Most of the known TNOs were discovered in `deep and narrow' observing campaigns using large
telescopes with small fields of view.  Current Near Earth Object (NEO) surveys \citep{Larson1998,
Kaiser2004} have the advantage of continuously monitoring large portions of the sky over several
years, but are disadvantaged because they use smaller telescopes with cadences designed to
identify NEOs that move more than $10\times$ faster than TNOs.  \citet{Brown2015} searched
archival data from the Catalina Sky Survey \citep{Larson2003} and Siding Spring Survey
\citep{Larson2003} and independently identified the eight brightest known TNOs.  Even though
they did not discover any new objects they predicted a 32\% chance that an object having
magnitude $V<19.1$ remains undiscovered in the unsurveyed region of the sky.


Evidence has been mounting in the past few years that there is a large planetary-sized distant
object in our solar system whose gravitational perturbations influence the orbits of
Scattered-Disc Objects (SDOs), particularly those on orbits similar to the dwarf planet (90377)
Sedna \citep[e.g.,][]{Trujillo2014,DeFuMarcos2015,Batygin2016}.  These works suggest that all
currently known extreme TNOs with semi-major axis greater than $150$~AU (including the only
other Sedna-like object: 2012~VP$_{113}$) show a pronounced clustering in their arguments of
perihelia ($\omega$) not present in the closer TNO population.  \citet{Trujillo2014} suggest
this clustering is centred at $\omega\sim0^{\circ}$ and that it is due to the Lidov---Kozai
effect, a three-body interaction capable of constraining $\omega$ \citep{Kozai1962}.  They
propose that a super-Earth mass body located at $\sim250$~AU would be capable of restricting
$\omega$ for these objects and be stable for billions of years.  However, \citet{Batygin2016}
made a similar calculation, but excluded orbits which do not demonstrate long term stability
because of Neptune, and found that the distant TNOs cluster around
$\omega\sim318^{\circ} \pm 8^{\circ}$ which is inconsistent with the Kozai mechanism.  They
suggest instead that the clustering can be maintained by a distant ten Earth-mass planet on
an eccentric orbit with semi-major axis $700$~AU, nearly co-planar with the distant TNOs,
but with $\omega$ shifted by $180^{\circ}$.  In addition, such a planet might explain the
presence of highly inclined TNOs whose existence has not yet been explained \citep{Gladman2009b}.
\citet{Trujillo2014} also state that another plausible explanation for such a peculiar asymmetric
$\omega$ configuration would be a strong stellar encounter with the Oort cloud in the past.

Increasing the number of known retrograde TNOs and Sedna-like SDOs is needed to further test
these hypotheses and to constrain the orbital elements and mass of any potentially undiscovered
planet.  Towards that end, in this work we report on the discovery and detection of the largest
number of TNOs by a single asteroid survey, which due to its long-duration and wide-field coverage,
provides an excellent complement to targeted deep-and-narrow surveys, resulting in a relatively
unbiased TNO sample.

\subsection{Pan-STARRS}

The prototype telescope for the Panoramic Survey Telescope and Rapid Response System (Pan-STARRS1,
hereafter referred to as PS1) located in the United States on Haleakala, Maui, Hawaii, has been
surveying the sky since 2010.  Many of the observations by PS1 were taken as a sequence of four
exposures, each separated by a Transient Time Interval (TTI) of $\sim20$ minutes.  This cadence
was selected to optimise detection of Near Earth Objects (NEOs) --- objects which have perihelia
$q<1.3$~AU.  Observations from each night are rapidly processed by the Image Processing Pipeline
\citep{Magnier2006}, and all detected moving objects identified by the Moving Object Processing
System \citep[MOPS;][]{Denneau2013} are reported to the Minor Planet Center.  PS1 has become the
leading discovery telescope for NEOs, discovering almost half of the new Near Earth Asteroids in
2015, and discovering more than half of the new comets in 2015 \citep{Wainscoat2015}.

The detection of NEOs is done using subtraction of image pairs which have a TTI of $\sim20$ minutes
and are well matched in image quality and telescope pointing.  This TTI spacing produces a lower limit
on the rate of motion for detection of moving objects, below which moving objects are self-subtracted
in their image pairs.  The lower limit is typically $\sim0.04^{\circ}$ per day ($=2\arcsec$ in $20$
~minutes), and is seeing dependent.  A substantial number of Centaurs (which we define as having
perihelia between Jupiter and Neptune) have been discovered from the pair-subtracted images, but
only a few more-distant objects have been reported from PS1 (see Figure \ref{real}), some of which
were discovered via the International Astronomical Search Collaboration
(IASC\footnote{http://iasc.hsutx.edu/}), an educational outreach program
in which images were blinked manually.

Now that PS1 has thoroughly surveyed the sky north of $-30^{\circ}$ declination, other methods
become viable for object detection that are potentially more sensitive to both fainter and slower
moving objects.  One method uses subtraction of a high-quality static sky image, derived from the
cumulative survey data.  The other method uses the historical survey to establish a catalogue of
stationary objects, and compares catalogues of new detections in new images to the static sky, to
reveal moving objects.

Over the course of the PS1 survey, image quality has improved, but the grid structure in the PS1 CCDs
requires many dithered images to produce a clean static sky image.  And although good images in the
\textit{gri} passbands are now available for much of the sky north of $-30^{\circ}$ declination,
the coverage in the more sensitive \textit{w} passband is more sparse, because surveying in that
band has been more focused on the ecliptic for the purpose of NEO discovery.  The PS1 survey has
also only recently been extended south to $-49^{\circ}$ declination.  For these reasons, we have
focussed our initial exploration of methods to extract fainter moving objects on the catalogue
based approach.

\section{Methodology}

To locate moving sources in the PS1 data, a new search method was developed and run on source
catalogues previously generated by the IPP \citep{Magnier2006} from PS1 images
taken between 2010 Feb 24 and 2015 July 31.  These catalogues are generated by a source extraction
program which identifies and measures the point spread function for objects in the images. A
detection is the information recorded about an object in a single exposure, and the catalogues
contain detections of moving objects as well as stationary sources which must be removed.

\begin{figure}[hp]
\centering
\includegraphics[width=0.9\textwidth]{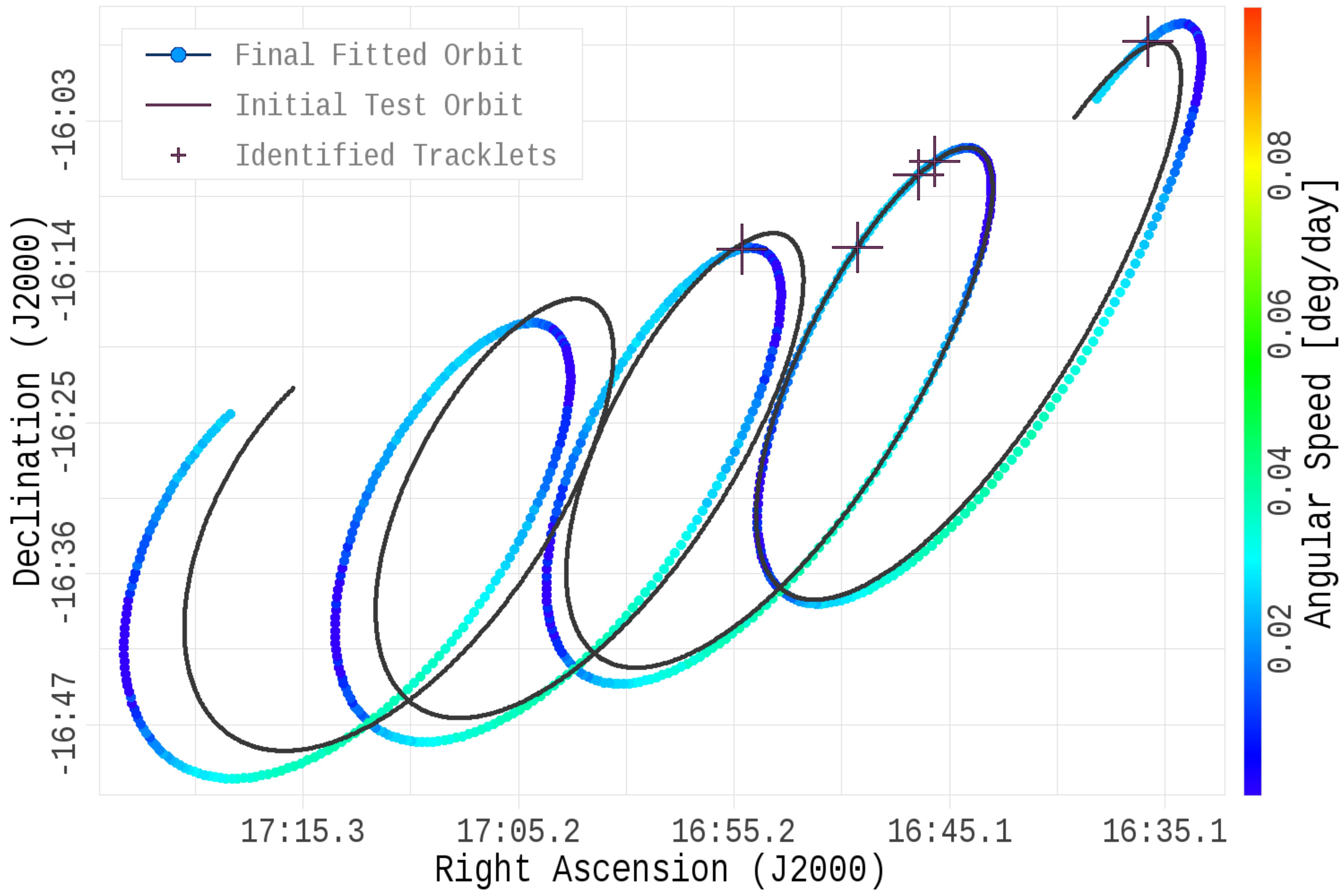}
\caption{The ephemerides of the initial orbit (determined from a tracklet pair) and the final
five-tracklet orbit for a sample event.  Because the motion cannot be represented with a linear
or quadratic projection (the loops are due to the motion of the Earth), a tracklet linking routine
based on ephemerides must be used.  For this event, the ephemerides from the two
orbits differ by $\sim9\arcmin$ after six months, and by $\sim15\arcmin$ after one year, which is
less than the $20\arcmin$ search radius described in the text.}
\label{path}
\end{figure}

The method, as described in the following sections, links sets of detections (corresponding to
the same object) from a single night into a `tracklet'.  Tracklets from multiple nights which
correspond to the same object are then searched for.  Figure \ref{path} shows the apparent path
across the celestial sphere over four years for a typical TNO.  Our method first searches for
two related tracklets which are used to generate an initial orbit from which ephemerides are
calculated to identify additional tracklets.  All detections identified for an object have
their image stamps extracted, which are visually inspected to ensure they are real and do
not correspond to image artefacts or stationary sources that were not removed.

While the PS1 MOPS pipeline \citep{Denneau2013} was designed to identify TNOs, it requires a survey
pattern and cadence that was not implemented.  The method presented here is able to link
TNOs using a survey pattern intended for NEO detection.

\subsection{Stationary Source Identification and Removal}

Objects in the outer solar system appear to move slowly across the celestial sphere, with
their apparent paths dominated by the motion of Earth.  TNOs in a 3:2 mean motion resonance
with Neptune, such as Pluto, will have a maximum angular speed at opposition of $0.022^{\circ}$
per day, equal to $3.2\arcsec$ per hour for orbits near zero inclination and eccentricity.
An object much farther out at $550$ AU will move $0.25\arcsec$ (the PS1 pixel scale) in one
hour.  However, even the most distant objects which move less than the astrometric uncertainty
in one night will move a noticeable amount over multiple nights, allowing visual verification
that they are indeed moving targets.

To identify stationary sources, consider a specific telescope pointing (defined by the boresight
direction and telescope rotation).  All source detections for all exposures overlapping the
field-of-view at this boresight direction are loaded.  For each detection, the number of
neighbouring detections within two magnitudes and $0.6\arcsec$ are counted.  Because each
PS1 telescope pointing is typically observed four times in a given night (the average TTI
is $19$~minutes, giving an average $57$~minute arc), we consider a detection stationary if
it is present more than four times, i.e., if there are detections at the same location on the
celestial sphere within $0.6\arcsec$ over multiple nights.  These criteria were determined
empirically and relate to the astrometric and photometric uncertainty in the PS1 images.

Once all stationary sources are identified, they are removed from the catalogues.  The detections
which remain are then used for tracklet creation.  Note that while valid detections may be rejected,
$100\%$ efficiency is not required as candidate objects are likely to have been observed multiple
times at different positions along their orbits.

\subsection{Tracklet Creation}

Tracklets are formed from detections following a procedure similar to the stationary source removal.
The detections from all exposures of a given pointing for a single night are loaded and iterated over.
All detections within $0.4$ magnitudes and $16\arcsec$
from each other are assigned a tracklet number.  The choice of $16\arcsec$ means many tracklets for
inner Solar System objects will be present in the tracklet dataset, as this corresponds to $0.43^
{\circ}$ per day for exposures with a TTI of $15$ minutes.  While these objects can be linked with
our algorithm, we exclude any identified object having semi-major axis $< 4.8$~AU, as we are only
interested in objects at heliocentric distances corresponding to Jupiter's orbit and beyond.  While
we could use a search distance $<16\arcsec$, this choice does allow us to identify tracklets where
the second or third detection occurred within a CCD cell gap.

Once formed, each tracklet has its motion along a great circle fit with a constant angular speed
model.  Only tracklets with RMS residual $< 0.3\arcsec$ and having $\ge3$ detections are kept,
and their fitted angular speed and position angle are recorded.  We do not create tracklets from
detection pairs because we cannot apply an RMS test to judge their astrometric quality.  Some
fraction of the formed tracklets may be faint stars or image artefacts which exhibit linear
motion, and their presence will increase the required computational time since there are more
tracklet comparisons to be made.  We validate all orbit fits using residual checks.


This linking process is different from the kd-tree based algorithm used by MOPS \citep{Denneau2013}
described in detail by \citet{Kubica2007}.  The method presented here is computationally faster,
but is limited to tracklets moving at much slower speed, and it cannot handle the intersection of
tracklets which correspond to detections of different objects with similar apparent magnitude.

\subsection{Tracklet Pairing}

\begin{figure}[p]
\centering
\includegraphics[width=0.9\textwidth]{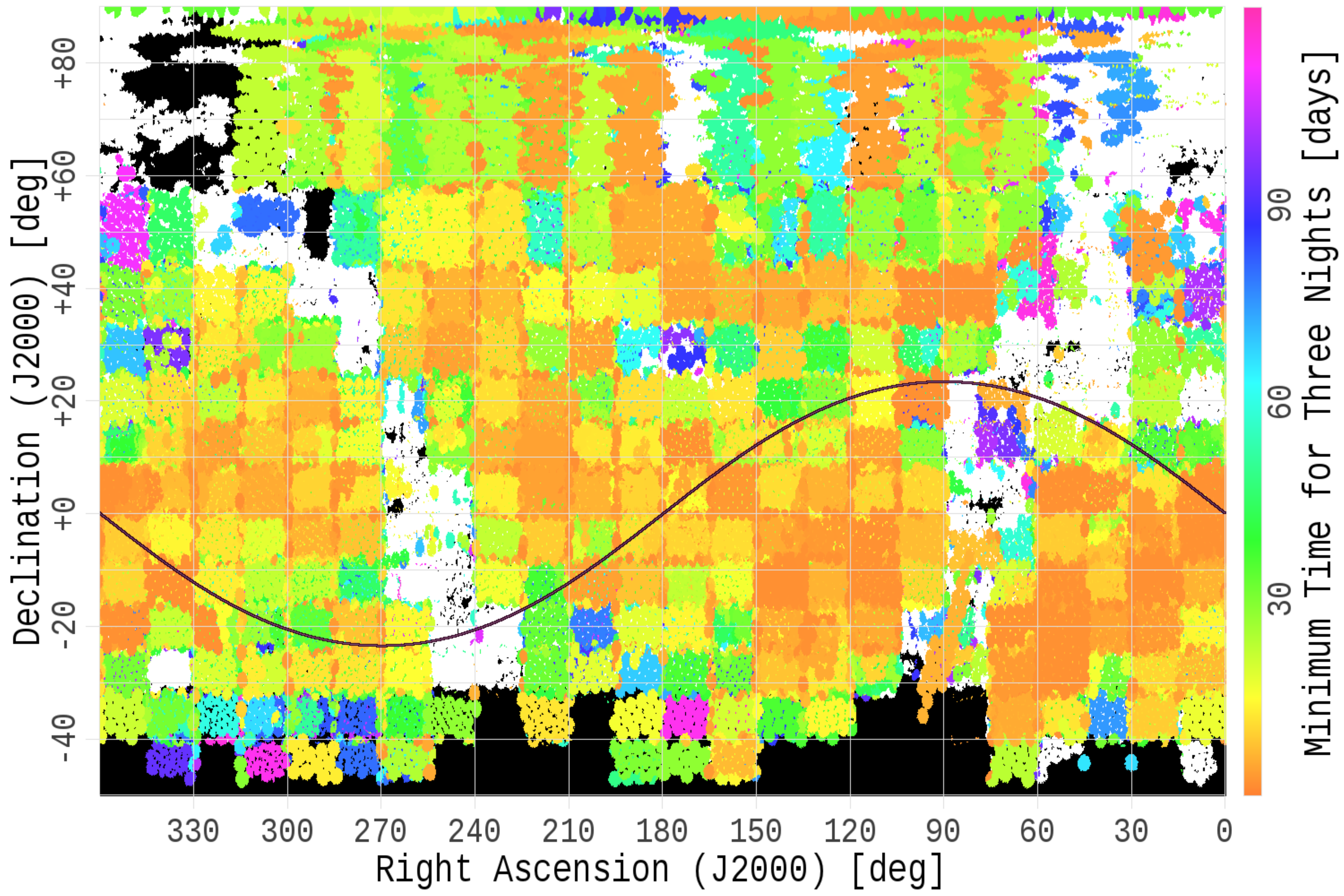}
\caption{Minimum timespan in the PS1 survey to observe the celestial sphere on three separate
nights, ignoring observations consisting of only one or two exposures.  The block pattern is
due to how the survey strategy was implemented.  The ecliptic plane is shown as a curved line.
 Black regions indicate three observations have never been made, while white (mostly corresponding
to the galactic plane) means there are more than $120$ days between the first and third nights.
The majority of the surveyed area satisfies the tracklet pairing search window of $60$ days.}
\label{look}
\end{figure}

To test if two tracklets correspond to the same object, a brute-force approach is used.  All
tracklet pairs occurring within up to $6^{\circ}$ (equal to the maximum angular speed of $0.1^{\circ}$
per day with a $60$ day window) have a test orbit fit using FindOrb
\footnote{http://www.projectpluto.com/find\_orb.htm}, and we
require its reported mean residual $<0.3\arcsec$.  The choice of a $60$ day window is a trade off
between an improved initial arc length, and the computation time needed to test all tracklet pairs,
and is reasonable since the PS1 survey pattern has observed much of the celestial sphere on at
least three separate nights within this timespan (see Figure \ref{look}).  However, this method
is affected by missed tracklets which may have fallen into CCD cell gaps, been obscured by other
image artefacts, or are too close to bright stars.  We do not require every tracklet pair for
an object to be identified, as many TNOs should be detected in more than two tracklets for the
timespan considered here (see Figure \ref{look}).  It is important to note however, that some
orbital geometries will not have any observations: for example, there are no PS1 observations
of the Centaur (10199) Chariklo which is located near the galactic centre.

To increase computational efficiency, tracklet pairs are identified using a kd-tree indexed
approach, and the position angle of both tracklets must be within $24^{\circ}$ of a line
connecting the two tracklets.  This value was empirically chosen and considerably
reduces the computation time required as it excludes certain unphysical geometries.  We
consider only tracklets having angular speed between $0.1^{\circ}$ and $0.001^{\circ}$ per
day, as our priority in this work is to find TNOs.  An extended search for more distant
objects will be considered in future work.

\subsection{Identifying additional tracklets}

The search for additional tracklets to extend an object's arc length requires a starting
orbit, for which we use the tracklet pairing results described in the previous section.
The metadata for each PS1 exposure is loaded, sorted by the observation time relative to
the starting orbit's epoch (where its mean anomaly is defined), and iterated over.  For
each exposure, an ephemeris is generated, and if located within the field-of-view, the
corresponding tracklet database is searched.  For each tracklet which contains detections
within $20\arcmin$ of the test ephemeris, a new orbit is fit using FindOrb.
If the reported mean residual is $<0.2\arcsec$, the tracklet is considered
linked, its astrometry appended, and the updated orbit fit kept.  The search then continues
for additional tracklets.  The choice of $0.2\arcsec$ is more strict than the limit used
during tracklet pairing, but has identified up to $27$ additional tracklets for the
candidate objects.

\begin{table}[ht]
\centering
\caption{Typical sample of a classical TNO (with $H=6.5$) showing how its orbit converges as
additional tracklets are found, with five total tracklets being identified.  All orbits are
integrated to the 20140702.0 epoch used for the initial two-tracklet case.  The first column
($N$) in the table is the number of tracklets used to compute the orbit.}
\begin{tabular}{cccccccc}
\\ [1ex]
\toprule
$N$ &  $a$ [AU]  & $e$ & $i$ [$^{\circ}$] & $\omega$ [$^{\circ}$] & $\Omega$ [$^{\circ}$] & $q$ [AU] & $M$ [$^{\circ}$] \\
\midrule
2 & 33.2933 & 0.0726  & 7.374 & 94.233 & 198.516 & 30.8756 & 325.536 \\
3 & 39.2198 & 0.2176  & 7.283 & 59.306 & 197.561 & 30.6844 & 357.708 \\
4 & 39.4907 & 0.2247  & 7.288 & 63.932 & 197.506 & 30.6186 & 354.920 \\
5 & 39.3916 & 0.2212  & 7.279 & 57.769 & 197.518 & 30.6778 & 358.711 \\
\bottomrule
\end{tabular}
\label{morb}
\end{table}

\begin{figure}[p]
\centering
\includegraphics[width=0.9\textwidth]{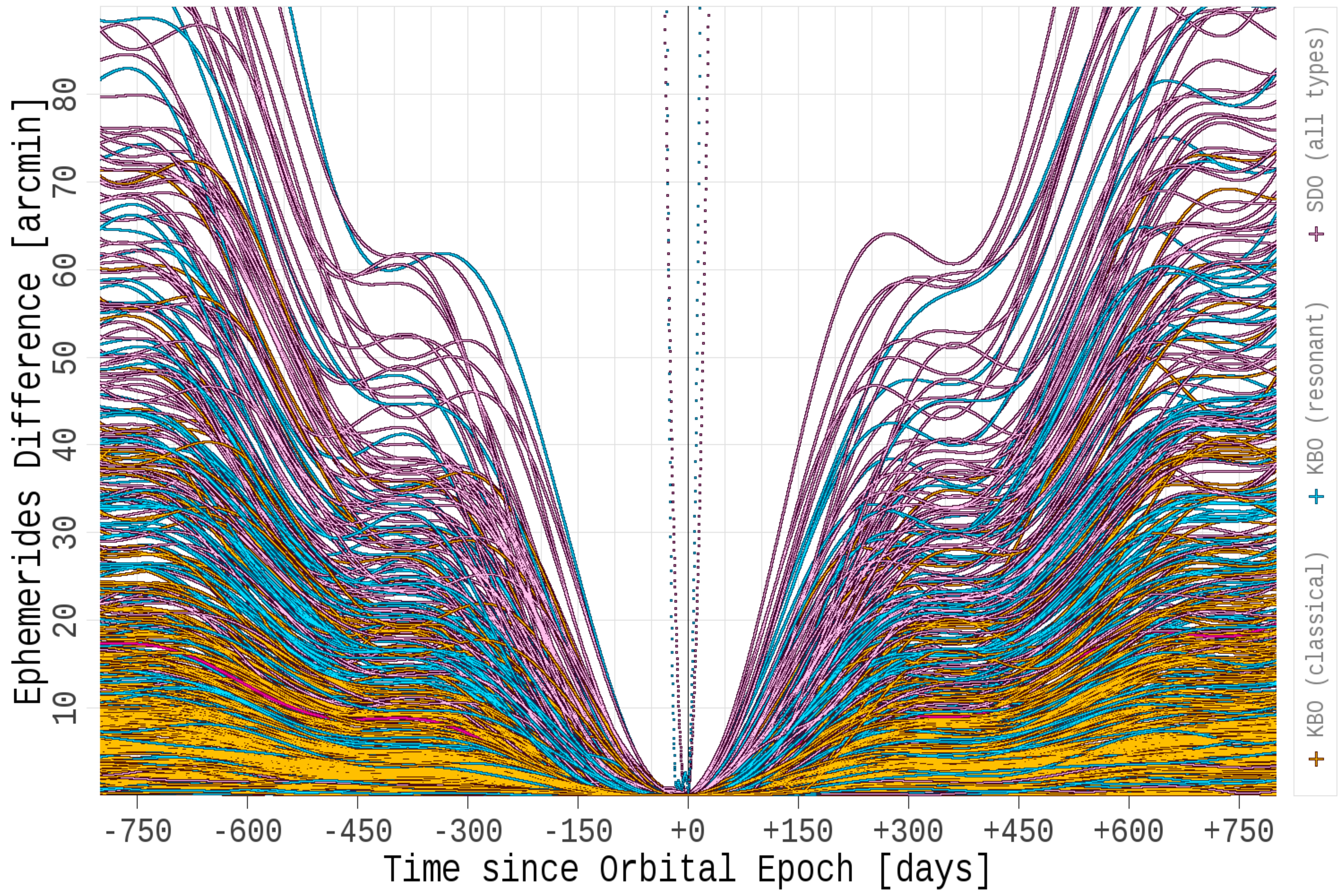}
\caption{The difference between the ephemeris from the initial tracklet-pair orbit and
the final orbit for all TNO events.  Using a tracklet search radius of $20\arcmin$
gives a $\sim250$ day window to find additional tracklets.  The two events reaching an
ephemeris error of $\sim90\arcmin$ within two weeks are due to particularly short arcs
in the initial tracklet pairing.} 
\label{slip}
\end{figure}

Table \ref{morb} shows how the initial test orbit of a sample object converges as
additional tracklets were found.  The difference in the ephemeris for the initial and
final orbit is illustrated in Figure \ref{path}.  The choice of the $20\arcmin$ search
distance was chosen based on identified TNOs: Figure \ref{slip} shows the angular
distance between ephemerides from the initial tracklet-pair based orbit and the final
fitted orbit for all identified objects.  The majority of events show $<20\arcmin$
difference over $\pm 120$~days, where (based on Figure \ref{look}) most of the celestial
sphere has been observed on at least three nights during the PS1 survey.

While we could integrate the test orbits using Mercury6 \citep{Chambers1999} to account
for their orbital evolution, this is not required for TNOs as the generated ephemerides
will change by much less than the $20\arcmin$ search distance.

\subsection{Visual Inspection and Classification}

The image stamps for each detection from all identified objects were extracted and visually
inspected to confirm they were real and not affected by image artefacts.

The orbital elements of the identified objects were used to classify them as potential Trojans,
classical or resonant TNOs, Scattered Disc Objects (SDOs), or Centaurs.  The orbital elements
were also used to identify known objects by comparing the location and position angle of all
ephemerides at their orbit epoch to those computed using the Minor Planet Center (MPC) orbit
catalogue after integration using Mercury6 \citep{Chambers1999}.  This method assumes the
orbits from the MPC are more accurate than those identified in this study, which may not be
true if our identified objects were independently previously observed by other telescopes
over shorter orbital arcs.

\section{Results and Discussion}
\label{results}

The source catalogues used in this study span the time from 2010 Feb 24 to 2015 July 31,
and are generated from $529\,609$ exposures in $61\,065$ pointings, with $93\,799\,429\,652$
total detections.  The stationary source removal left $7\,655\,731\,998$ detections of which
$232\,447\,038$ were linked into $65\,524\,472$ tracklets.

\begin{figure}[p]
\centering
\includegraphics[width=0.9\textwidth]{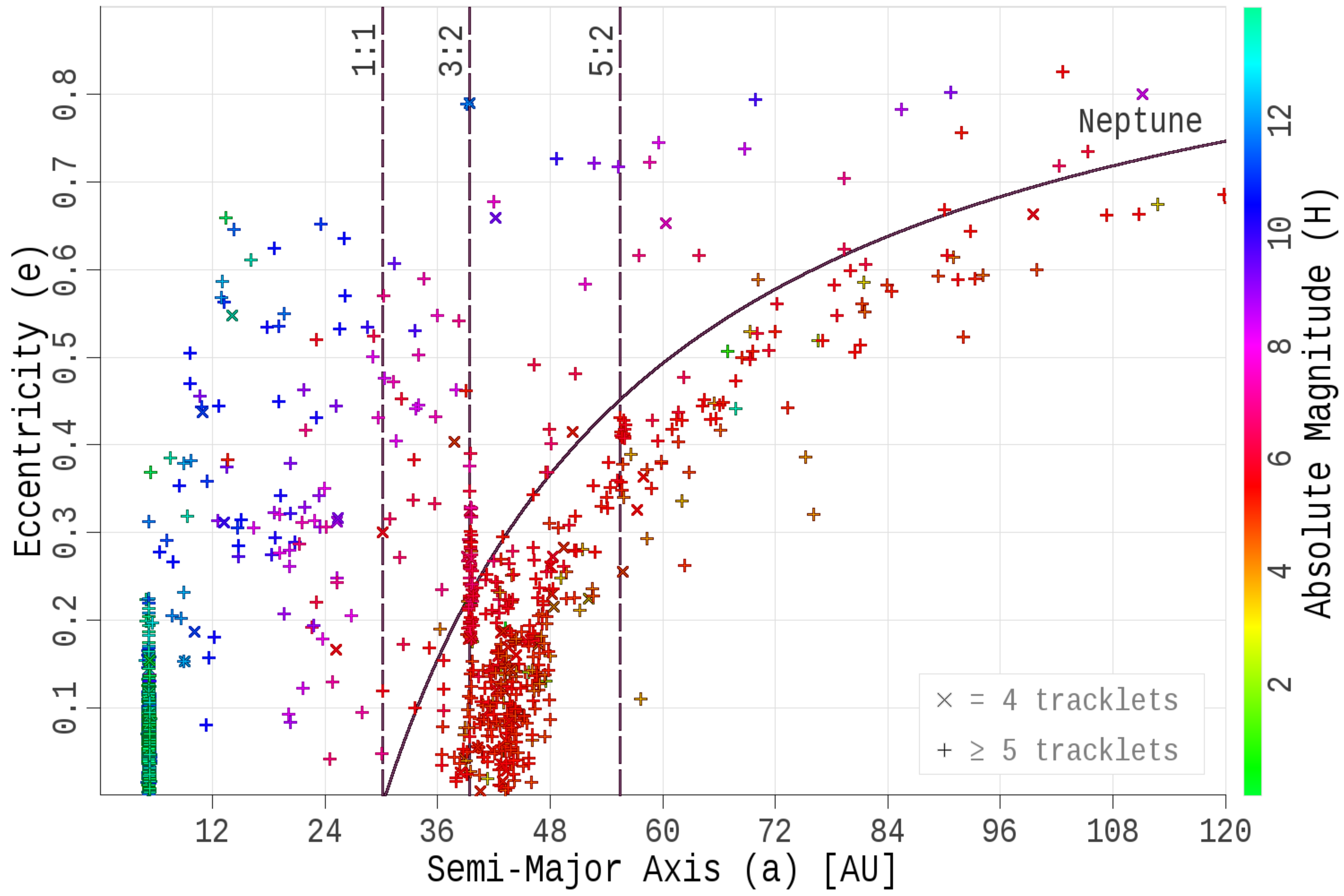}
\caption{Eccentricity versus semi-major axis for the objects identified in this work,
colour coded by absolute magnitude.  The detected objects consist of $789$ Jupiter Trojans,
$154$ Centaurs, and $465$ total TNOs.  The $1:1$ and $3:2$ mean motion resonances with Neptune
are noted with vertical dashed lines, while Neptune's aphelion distance of $q=30.4$~AU is noted
with a solid curved line.}
\label{figA}
\end{figure}

Figure \ref{figA} shows eccentricity vs semi-major axis for all identified objects,
and we present a breakdown of these in Table \ref{count}, including the number of unknown
and known objects, as well as the number expected (both total and to a limiting $V=22.5$)
based on the MPC catalogue.  We further discuss our identified objects in the following
sections.

\begin{table}[ht]
\centering
\caption{Classification of identified objects, also broken down into known and unknown  objects.
The fifth column shows the expected number from the MPC catalogue as a fraction of the total for
a limiting magnitude of $V=22.5$.}
\begin{tabular}{lccccc}
\\ [1ex]
\toprule
orbital type   & total   & unknown & known & MPC ($V<22.5$) & MPC total \\
\midrule
 Jupiter Trojans	 &  $789$ &  145 & $644$ & $6169$ & $6207$ \\
 Centaurs		 &  $154$ &   78 &  $76$ &  $211$ &  $303$ \\
 Classical TNOs		 &  $255$ &  162 &  $91$ &  $197$ & $1198$ \\
 Resonant TNOs		 &  $121$ &   77 &  $44$ &   $93$ &  $319$ \\
 Scattered Disc		 &   $89$ &   52 &  $37$ &   $57$ &  $203$ \\
\bottomrule
\end{tabular}
\label{count}
\end{table}

\subsection{Bright Objects}

\citet{Brown2015} estimate that there is a 32\% chance that a TNO having magnitude $V<19.1$ remains
undiscovered after their archival search of the Catalina Sky Survey database.  We computed ephemerides
for the five year period of data used in this study, for all identified objects, and list the peak
brightness of the top eight in Table \ref{bright}.  There is one object which reaches $V=18.5$, which
is the third brightest TNO over the timespan of data used in this study.

\begin{table}[ht]
\centering
\caption{The brightest eight objects (between 2010 Feb 24 and 2015 July 31) identified
in our study.  The first column gives the total tracklet count for each object.  The
magnitude column ($V_{min}$) gives the brightest apparent $V$ magnitude that each object
reached during the five year PS1 survey data used based on the absolute magnitude ($H$)
determined from FindOrb, and $N$ is the number of tracklets used for
the orbit computation.}
\begin{tabular}{ccccccl}
\\ [1ex]
\toprule
$N$ & $a$ [AU]  & $e$ & $i$ [$^{\circ}$] & $H$ & $V_{min}$ & object \\
\midrule
13 & 43.2588 & 0.1905 & 28.193 &  0.1 & 17.2 & (136108) Haumea		\\ 
11 & 67.7662 & 0.4410 & 44.057 & -1.6 & 18.2 & (136199) Eris		\\ 
11 & 36.2522 & 0.1894 &  1.497 &  3.8 & 18.5 & \textbf{unknown}		\\ 
12 & 39.4661 & 0.2182 & 20.554 &  1.8 & 18.6 & (90482) Orcus		\\ 
16 & 39.5548 & 0.2786 & 15.464 &  4.1 & 18.6 & (38628) Huya		\\ 
25 & 39.3913 & 0.2243 &  8.411 &  4.2 & 19.0 & (47171) 1999 TC$_{36}$	\\ 
06 & 69.2813 & 0.5289 & 29.424 &  3.3 & 19.0 & 2010 EK$_{139}$		\\ 
16 & 41.3601 & 0.0193 & 19.307 &  3.1 & 19.2 & (145452) 2005 RN$_{43}$	\\ 
\bottomrule
\end{tabular}
\label{bright}
\end{table}

\subsection{Large Objects}

\begin{figure}[p]
\centering
\includegraphics[width=0.9\textwidth]{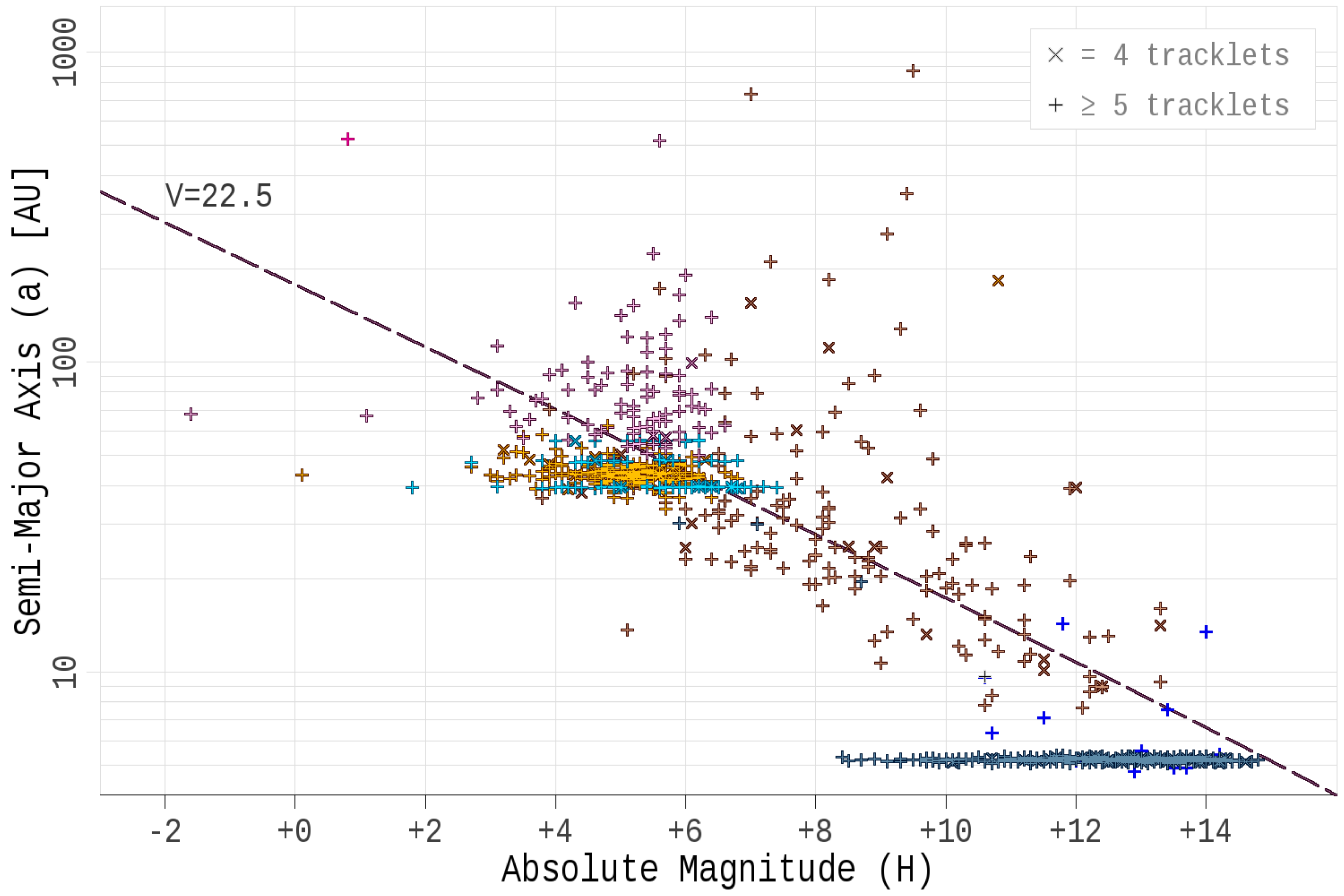}
\caption{Semi-major axis vs absolute magnitude for all identified objects.  The
cluster along $a\sim5.2$\,AU at the lower right are Jupiter Trojans.  The largest
objects (left most in the figure) are all known.}
\label{figC}
\end{figure}

Figure \ref{figC} plots semi-major axis vs absolute magnitude for all identified
objects.  The largest identified objects are all known, and are further listed
in Table \ref{biggies}.  However, (136472) Makemake and (134340) Pluto are not
identified.  For the first case, its tracklets do not occur within a $60$ day
window, and for the latter, the only tracklet pair occurred with a two day arc,
and additional tracklets were not linked.


There are no unknown distant planetary-sized objects identified in our study.  However,
our minimum tracklet angular speed cut-off of $0.001^{\circ}$ per day could exclude them
if they are present in the source catalogues.

\begin{table}[ht]
\centering
\caption{The most distant objects identified in our study.  The column headings are
identical to those given in Figure \ref{bright}.}
\begin{tabular}{ccccccl}
\\ [1ex]
\toprule
$N$ &  $a$ [AU]  & $e$ & $i$ [$^{\circ}$] & $H$ & $V_{min}$ & object \\
\midrule
11 &  67.7662 & 0.4410 & 44.057 & -1.60 & 18.2 & (136199) Eris			\\ 
13 &  43.2588 & 0.1905 & 28.193 &  0.10 & 17.2 & (136108) Haumea		\\ 
12 & 525.8952 & 0.8554 & 11.929 &  0.80 & 20.1 & (90377) Sedna			\\ 
19 &  66.8888 & 0.5065 & 30.946 &  1.10 & 20.4 & (225088) 2007 OR$_{10}$	\\ 
12 &  39.4661 & 0.2182 & 20.554 &  1.80 & 18.6 & (90482) Orcus			\\ 
18 &  45.7416 & 0.1402 & 21.497 &  2.70 & 19.4 & (174567) Varda			\\ 
11 &  47.5112 & 0.1304 & 24.337 &  2.70 & 19.3 & (55565) 2002 AW$_{197}$	\\ 
11 &  76.5622 & 0.5186 & 23.344 &  2.80 & 19.2 & (229762) 2007 UK$_{126}$	\\ 
13 &  43.1700 & 0.0506 & 17.154 &  3.00 & 19.4 & (20000) Varuna			\\ 
10 & 112.7967 & 0.6744 & 14.025 &  3.10 & 20.0 & (303775) 2005 QU$_{182}$	\\ 
10 &  81.4754 & 0.5860 &  7.565 &  3.10 & 21.2 & 2015 RR$_{245}$		\\ 
\midrule
\bottomrule
\end{tabular}
\label{biggies}
\end{table}

\subsection{Distant Objects}

\begin{figure}[p]
\centering
\includegraphics[width=0.9\textwidth]{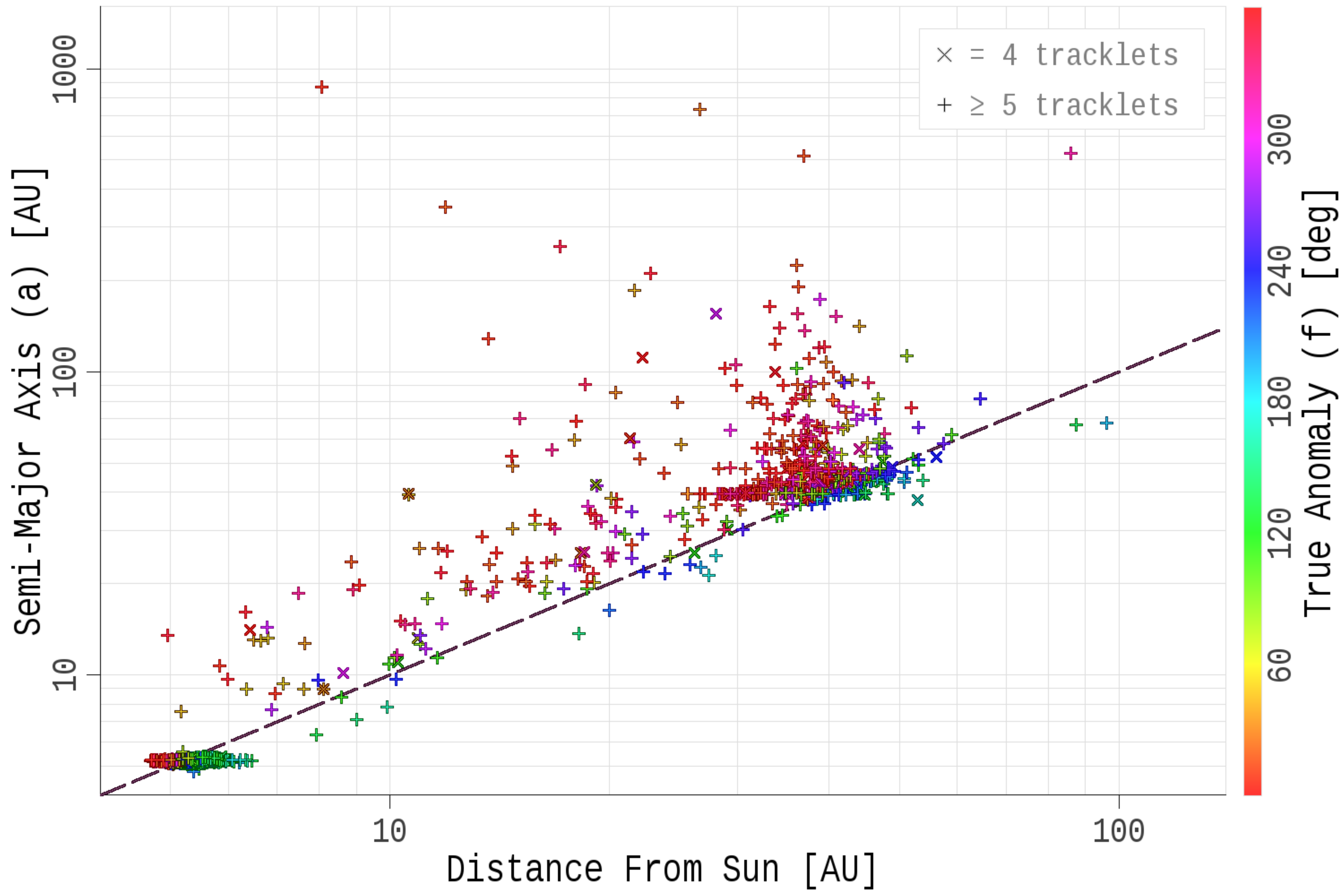}
\caption{Semi-major axis vs distance from the Sun (on 2015 July 31) for all identified
objects.  The colour scheme shows the true anomaly, which gives a direct measure of
where each object is along its orbit relative to perihelion.  Objects significantly
above the dashed line have much larger eccentricity, with comets being located towards
the upper left of the plot.}
\label{figE}
\end{figure}

Figure \ref{figE} plots semi-major axis vs distance from the Sun (on 2015 July 31).
The farthest of these objects are listed in Table \ref{distant}, with the
six most distant being known (including the recent 2015 RR$_{245}$
\footnote{http://www.minorplanetcenter.net/mpec/K16/K16N67.html}).  In addition to these,
we have identified two previously unknown objects which have perihelia well beyond the
orbit of Neptune.

\begin{table}[ht]
\centering
\caption{The most distant identified objects in our dataset, where $r_{max}$ is their
heliocentric distance on 2015 July 31.  The first column ($N$) gives the tracklet count
for each object.  Two of the identified objects are not known.}
\begin{tabular}{ccccccl}
\\ [1ex]
\toprule
$N$  &  $a$ [AU]  & $e$ & $i$ [$^{\circ}$] & $H$ & $r_{max}$ & object \\
\midrule
11 & 67.7662 & 0.4410 & 44.057 & -1.6 & 96.3 & (136199) Eris		\\ 
19 & 66.8888 & 0.5065 & 30.946 & 1.1 & 87.3 & (225088) 2007 OR$_{10}$	\\ 
12 &525.8952 & 0.8554 & 11.929 & 0.8 & 85.9 & (90377) Sedna		\\ 
10 & 81.4754 & 0.5860 & 7.565 & 3.1 &  64.5 & 2015 RR$_{245}$		\\ 
14 & 62.0774 & 0.3355 & 28.814 & 3.4 & 58.9 & 2015 KH$_{162}$		\\ 
07 & 57.7016 & 0.1103 & 46.604 & 3.5 & 57.5 & 2004 XR$_{190}$		\\ 
19 & 62.1612 & 0.2245 & 10.252 & 3.2 & 56.2 & \textbf{unknown}		\\ 
08 & 43.9118 & 0.2511 & 14.878 & 4.0 & 53.8 & \textbf{unknown}		\\ 
\bottomrule
\end{tabular}
\label{distant}
\end{table}

\subsection{Orbital Clustering}

\begin{figure}[p]
\centering
\includegraphics[width=0.9\textwidth]{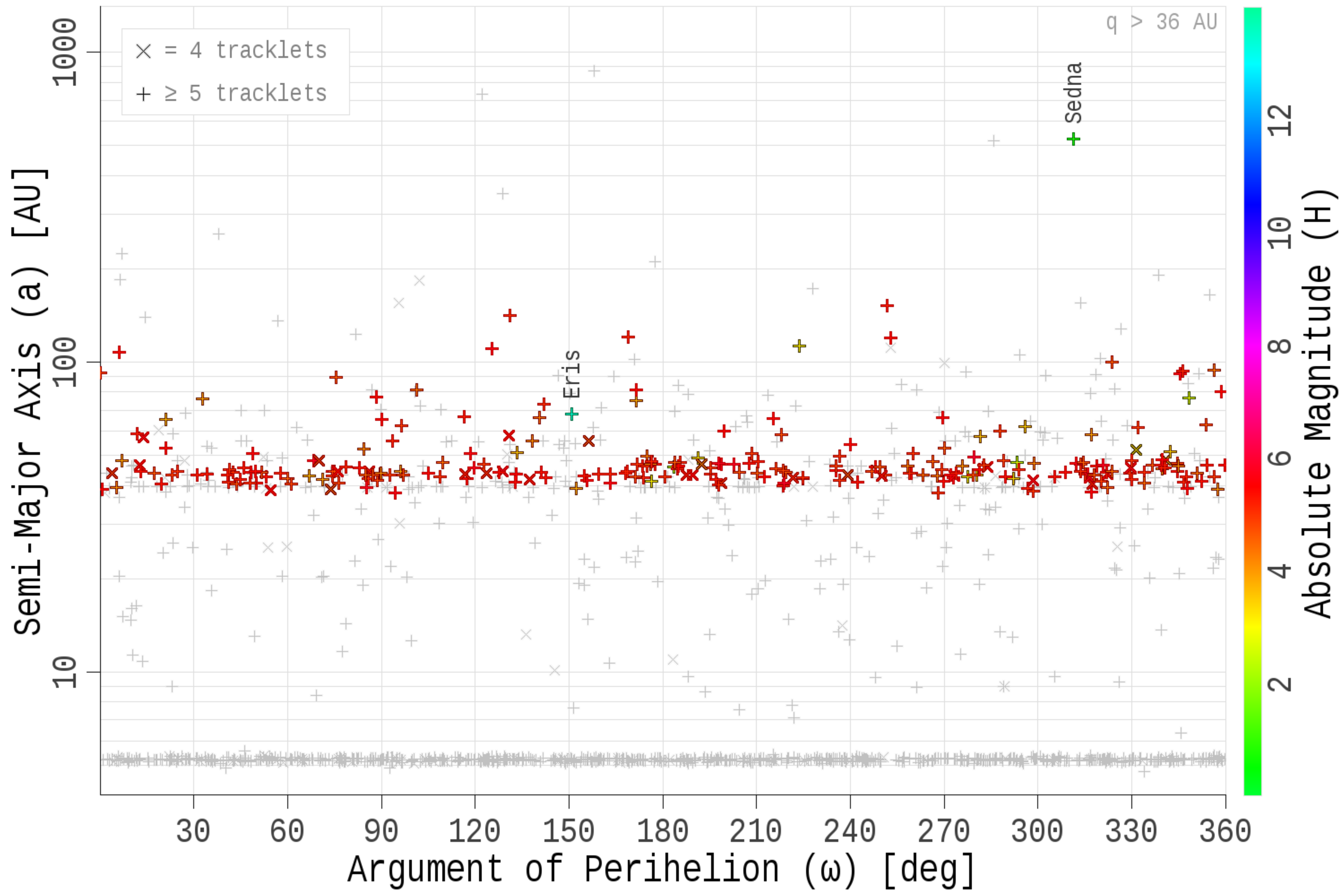}
\caption{Semi-major axis versus argument of perihelion for all identified objects having
$q >36$~AU.  The location of (90377) Sedna and (136199) Eris are indicated.  Sedna is the
only object identified in this survey that supports the orbital clustering discussed in
\citet{Trujillo2014} and \citet{Batygin2016}.}
\label{figB}
\end{figure}

\citet{Batygin2016} suggest the orbital configuration of TNOs decoupled from Neptune's
influence support the presence of a planetary-sized perturber in the outer solar system.
Due to the sensitivity limit of PS1 ($V\sim22.5$), we can only identify two of these
objects, namely (90377) Sedna and 2007 TG$_{422}$, and both were identified in this study.
We present the arguments of perihelia of our identified objects in Figure \ref{figB}.
Using a Kolmogorov-Smirnov test on all identified objects having $q > 32$~AU and
$q > 36$~AU, we do not see clustering in $\omega$.  To our sensitivity limit, this
is significant as our wide field survey might be expected to have less observational
bias than targeted narrow searches.  Survey bias might also be expected to induce apparent
clustering in an observed population, rather than remove it.

\subsection{High Inclination Objects}

\begin{figure}[p]
\centering
\includegraphics[width=0.9\textwidth]{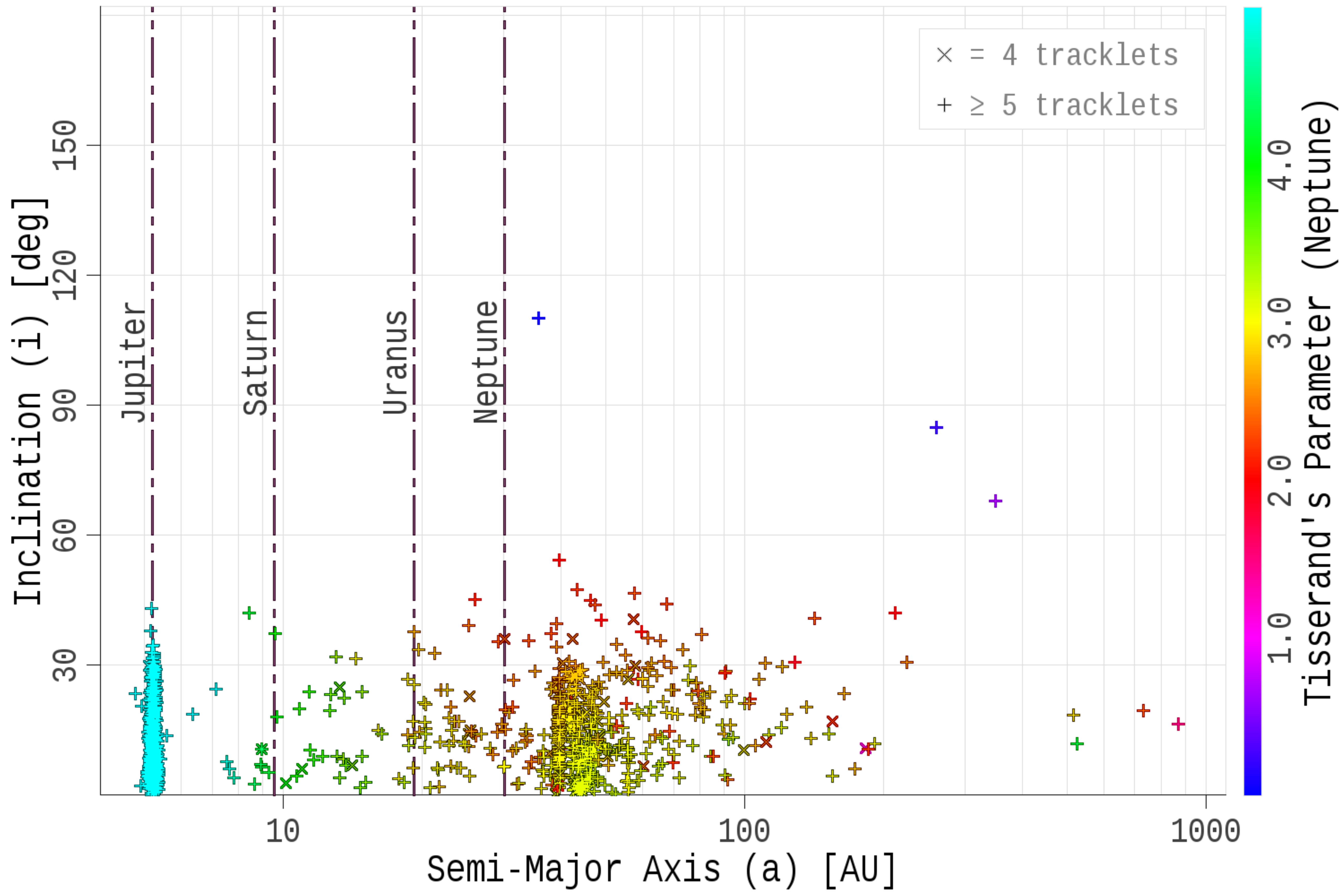}
\caption{Inclination vs semi-major axis for all identified objects, colour coded by Tisserand's
parameter wrt. Neptune.  The location of the giant planets are noted with vertical lines, with
a large grouping of Jupiter Trojans towards the lower left.  The identified object with highest
eccentricity is known object 2011 KT$_{19}$.}
\label{figD}
\end{figure}

From Figure \ref{figD}, there are several identified objects with high inclination, and we
list those having the largest inclination in Table \ref{retro}.  We identified in our study
the recent 2011 KT$_{19}$ \footnote{Minor Planet Circulars Orbit Supplement 380701}, and
also have one unknown object with $i=84^{\circ}$.  The orbital evolution of highly-inclined
TNOs is not understood, but \citet{Gladman2009b} suggest they could have been perturbed into
their current orbits by a planetary-sized perturber in the outer solar system.  Another
plausible explanation is that they are captured objects.

\begin{table}[ht]
\centering
\caption{The identified objects (excluding comets) with highest inclination.  The
magnitude ($V_{min}$) is the brightest each objects reaches over the five year data
span used, and is determined from the absolute magnitude ($H$) as fitted by FindOrb.
The first column ($N$) gives the number of tracklets used in the orbit computation.}
\begin{tabular}{ccccccl}
\\ [1ex]
\toprule
$N$  &  $a$ [AU]  & $e$ & $i$ [$^{\circ}$] & $H$ & $V_{min}$ & object \\
\midrule
19 &  35.6490 & 0.3330 & 110.206 &  6.6 & 20.5 & 2011 KT$_{19}$			\\ 
06 & 259.7509 & 0.9354 &  84.824 &  9.1 & 21.5 &  \textbf{unknown}		\\ 
06 & 349.2179 & 0.9685 &  68.030 &  9.4 & 20.0 & (418993) 2009 MS$_{9}$		\\ 
06 &  39.4982 & 0.3899 &  54.265 &  6.5 & 20.6 & \textbf{unknown}		\\ 
10 &  43.3078 & 0.2160 &  47.507 &  6.0 & 21.3 & \textbf{unknown}		\\ 
07 &  57.7016 & 0.1103 &  46.604 &  3.5 & 21.1 & 2004 XR$_{190}$		\\ 
\bottomrule
\end{tabular}
\label{retro}
\end{table}

\subsection{Neptune Trojans}

We list in Table \ref{trojs} two possible Neptune Trojans present in our identified
objects, based solely on their orbital elements.  One of these objects was identified
in a parallel study using PS1 data \citep{2015DPS....4721117L}, however we did not
identify their other candidates.  We present a discussion as to why this might be in
Section \ref{whyyyy}.

\begin{table}[ht]
\centering
\caption{Two possible Neptune Trojans (based on their orbital elements) identified.
The first column ($N$)indicates the number of tracklets used to compute the orbit.}
\begin{tabular}{ccccccl}
\\ [1ex]
\toprule
$N$  &  $a$ [AU]  & $e$ & $i$ [$^{\circ}$] & $H$ & $V_{min}$ & object \\
\midrule
16 & 30.1564 & 0.1192 &  6.661 & 5.90 & 20.7 & 2013 KY$_{18}$	\\ 
06 & 30.0280 & 0.0472 &  6.562 & 7.10 & 21.6 & 2010 TS$_{191}$	\\ 
\bottomrule
\end{tabular}
\label{trojs}
\end{table}

\subsection{Centaurs and Comets}

We define Centaurs to have perihelia between Jupiter and Neptune, but which are not in a mean
motion resonance with Neptune.  We identified $136$ Centaurs in our study, as well as $10$ comets
which were all known objects.  Our identification method does not link objects having $e\ge1$.
Although \citet{2013AJ....146...36S} call for a better characterised survey of Centaur objects,
we consider them a by-product of the current study and leave this to a future publication ---
our primary interest here is identifying TNOs.

%

\subsection{Identification Efficiency}
\label{whyyyy}

Our discussion would not be complete without mentioning the efficiency of creating and pairing
tracklets, as well as identifying additional tracklets.  As suggested by Table \ref{count},
our method identifies about half of the expected (i.e., having $V<22.5$) known population of
classical and resonant TNOs from the MPC catalogue, and two-thirds of the known SDOs.  We do
not count Jupiter Trojans or Centaur objects here because our detection parameters were
empirically chosen to optimise detection of TNOs.

A complete end-to-end measure of the identification efficiency could be made by injecting
synthetic detections into our software routines \citep[e.g., as is done by MOPS;][]{Denneau2013},
but ideally this must be done directly into the original PS1 image data which would then be used
to produce new source catalogues.  This reprocessing by the IPP \citep{Magnier2006} is necessary
for a realistic efficiency determination as the fill-factor of the PS1 camera is limited to
$\sim70\%$, and our tracklet linking routines have no knowledge of which regions on the CCD are
masked.  Also, our limiting sensitivity is dependent on the photometric passband as well as local
weather conditions.  Because of these reasons, determining an accurate efficiency is a significant
undertaking, which we will address in subsequent work to obtain an unbiased estimate of the true
TNO population.

From the current MPC catalogue for all known objects, the average distance between each asteroids
within $30^{\circ}$ of the ecliptic and its closest neighbour having apparent magnitude within
$0.4$ is $\sim820\arcsec$.  This is much greater than the $16\arcsec$ search radius used during
the tracklet creation process, suggesting that the majority of tracklets in the source catalogue
(after removing stationary sources) will not be contaminated by detections associated with different
objects.  Our tracklets may be limited in terms of their RMS residual, but even for a distant object
at $100$~AU, detections will be spaced $0.45\arcsec$ for a $20$ minute TTI, which is twice the PS1
pixel scale.

Because Figure \ref{slip} suggests additional tracklets should be readily found, this implies
the efficiency of our search implementation is limited by either the stationary source removal
process, or the tracklet pairing stage.  We also note that while the PS1 camera fill-factor may
lead to missed detections for NEOs, it can result in completely missed tracklets for TNOs due
to their much slower speed across the celestial sphere.

\section{Conclusions}

A search for distant objects was made using the archival PS1 data, with $1420$ objects identified,
consisting of $255$ classical TNOs, $121$ resonant TNOs, $89$ SDOs, $154$ Centaurs, and $789$
Jupiter Trojans.  Excluding the trojans, $371$ of these are new objects which we could not link
to known objects.

While our identified objects do not show a clustering in their arguments of perihelia, increasing
the number of known retrograde TNOs and Sedna-like SDOs is important in better understanding
the distant population in our solar system, especially to constrain the orbital elements and mass
of any potentially undiscovered large planetary-sized objects \citep{Trujillo2014,Batygin2016}.

Future work will focus on validating the detection efficiency, as well as optimising our
detection parameters to work well beyond the classical TNO regime.

\acknowledgements
\section*{Acknowledgments}

The Pan-STARRS1 Surveys (PS1) have been made possible through contributions of the Institute for
Astronomy, the University of Hawaii, the Pan-STARRS Project Office, the Max-Planck Society and its
participating institutes, the Max Planck Institute for Astronomy, Heidelberg and the Max Planck
Institute for Extraterrestrial Physics, Garching, The Johns Hopkins University, Durham University,
the University of Edinburgh, Queen's University Belfast, the Harvard-Smithsonian Center for Astrophysics,
the Las Cumbres Observatory Global Telescope Network Incorporated, the National Central University of
Taiwan, the Space Telescope Science Institute, the National Aeronautics and Space Administration under
Grant Nos. NNX08AR22G, NNX12AR65G, and NNX14AM74G issued through the Planetary Science Division of the
NASA Science Mission Directorate, the National Science Foundation under Grant No. AST-1238877, the
University of Maryland, and Eotvos Lorand University (ELTE) and the Los Alamos National Laboratory.

\bibliographystyle{aasjournal}
\bibliography{ps1tno}

\begin{thebibliography}{}
\expandafter\ifx\csname natexlab\endcsname\relax\def\natexlab#1{#1}\fi

\bibitem[{{Alexandersen} {et~al.}(2014){Alexandersen}, {Gladman}, {Kavelaars},
  {Petit}, {Gwyn}, \& {Shankman}}]{Alexandersen2014}
{Alexandersen}, M., {Gladman}, B., {Kavelaars}, J.~J., {et~al.} 2014, ArXiv
  e-prints, arXiv:1411.7953

\bibitem[{{Batygin} \& {Brown}(2016)}]{Batygin2016}
{Batygin}, K., \& {Brown}, M.~E. 2016, \aj, 151, 22

\bibitem[{{Bernstein} {et~al.}(2004){Bernstein}, {Trilling}, {Allen}, {Brown},
  {Holman}, \& {Malhotra}}]{Bernstein2004}
{Bernstein}, G.~M., {Trilling}, D.~E., {Allen}, R.~L., {et~al.} 2004, \aj, 128,
  1364

\bibitem[{{Brown} {et~al.}(2015){Brown}, {Bannister}, {Schmidt}, {Drake},
  {Djorgovski}, {Graham}, {Mahabal}, {Donalek}, {Larson}, {Christensen},
  {Beshore}, \& {McNaught}}]{Brown2015}
{Brown}, M.~E., {Bannister}, M.~E., {Schmidt}, B.~P., {et~al.} 2015, ArXiv
  e-prints, arXiv:1501.00941

\bibitem[{{Chambers}(1999)}]{Chambers1999}
{Chambers}, J. 1999, MNRAS, 304, 793

\bibitem[{{de la Fuente Marcos} {et~al.}(2015){de la Fuente Marcos}, {de la
  Fuente Marcos}, \& {Aarseth}}]{DeFuMarcos2015}
{de la Fuente Marcos}, C., {de la Fuente Marcos}, R., \& {Aarseth}, S.~J. 2015,
  \mnras, 446, 1867

\bibitem[{{Denneau} {et~al.}(2013){Denneau}, {Jedicke}, {Grav}, {Granvik},
  {Kubica}, {Milani}, {Vere{\v s}}, {Wainscoat}, {Chang}, {Pierfederici},
  {Kaiser}, {Chambers}, {Heasley}, {Magnier}, {Price}, {Myers}, {Kleyna},
  {Hsieh}, {Farnocchia}, {Waters}, {Sweeney}, {Green}, {Bolin}, {Burgett},
  {Morgan}, {Tonry}, {Hodapp}, {Chastel}, {Chesley}, {Fitzsimmons}, {Holman},
  {Spahr}, {Tholen}, {Williams}, {Abe}, {Armstrong}, {Bressi}, {Holmes},
  {Lister}, {McMillan}, {Micheli}, {Ryan}, {Ryan}, \& {Scotti}}]{Denneau2013}
{Denneau}, L., {Jedicke}, R., {Grav}, T., {et~al.} 2013, \pasp, 125, 357

\bibitem[{{Elliot} {et~al.}(2005){Elliot}, {Kern}, {Clancy}, {Gulbis},
  {Millis}, {Buie}, {Wasserman}, {Chiang}, {Jordan}, {Trilling}, \&
  {Meech}}]{Elliot2005}
{Elliot}, J.~L., {Kern}, S.~D., {Clancy}, K.~B., {et~al.} 2005, \aj, 129, 1117

\bibitem[{{Fraser}(2009)}]{Fraser2009}
{Fraser}, W.~C. 2009, \apj, 706, 119

\bibitem[{{Fuentes} \& {Holman}(2008)}]{Fuentes2008}
{Fuentes}, C.~I., \& {Holman}, M.~J. 2008, \aj, 136, 83

\bibitem[{{Gladman} {et~al.}(2001){Gladman}, {Kavelaars}, {Petit},
  {Morbidelli}, {Holman}, \& {Loredo}}]{Gladman2001}
{Gladman}, B., {Kavelaars}, J.~J., {Petit}, J.-M., {et~al.} 2001, \aj, 122,
  1051

\bibitem[{{Gladman} {et~al.}(2009){Gladman}, {Kavelaars}, {Petit}, {Ashby},
  {Parker}, {Coffey}, {Jones}, {Rousselot}, \& {Mousis}}]{Gladman2009b}
{Gladman}, B., {Kavelaars}, J., {Petit}, J.-M., {et~al.} 2009, \apjl, 697, L91

\bibitem[{{Gladman} {et~al.}(2012){Gladman}, {Lawler}, {Petit}, {Kavelaars},
  {Jones}, {Parker}, {Van Laerhoven}, {Nicholson}, {Rousselot}, {Bieryla}, \&
  {Ashby}}]{Gladman2012}
{Gladman}, B., {Lawler}, S.~M., {Petit}, J.-M., {et~al.} 2012, \aj, 144, 23

\bibitem[{{Jewitt} \& {Luu}(1993)}]{Luu1993}
{Jewitt}, D., \& {Luu}, J. 1993, Nature, 362, 730

\bibitem[{{Kaiser}(2004)}]{Kaiser2004}
{Kaiser}, N. 2004, in Society of Photo-Optical Instrumentation Engineers (SPIE)
  Conference Series, Vol. 5489, Society of Photo-Optical Instrumentation
  Engineers (SPIE) Conference Series, ed. J.~M. {Oschmann}, Jr., 11--22

\bibitem[{{Kozai}(1962)}]{Kozai1962}
{Kozai}, Y. 1962, \aj, 67, 591

\bibitem[{{Kubica} {et~al.}(2007){Kubica}, {Denneau}, {Grav}, {Heasley},
  {Jedicke}, {Masiero}, {Milani}, {Moore}, {Tholen}, \&
  {Wainscoat}}]{Kubica2007}
{Kubica}, J., {Denneau}, L., {Grav}, T., {et~al.} 2007, \icarus, 189, 151

\bibitem[{{Larsen} {et~al.}(2001){Larsen}, {Gleason}, {Danzl}, {Descour},
  {McMillan}, {Gehrels}, {Jedicke}, {Montani}, \& {Scotti}}]{Larsen2001}
{Larsen}, J.~A., {Gleason}, A.~E., {Danzl}, N.~M., {et~al.} 2001, \aj, 121, 562

\bibitem[{{Larson} {et~al.}(2003){Larson}, {Beshore}, {Hill}, {Christensen},
  {McLean}, {Kolar}, {McNaught}, \& {Garradd}}]{Larson2003}
{Larson}, S., {Beshore}, E., {Hill}, R., {et~al.} 2003, in Bulletin of the
  American Astronomical Society, Vol.~35, AAS/Division for Planetary Sciences
  Meeting Abstracts \#35, 982

\bibitem[{{Larson} {et~al.}(1998){Larson}, {Brownlee}, {Hergenrother}, \&
  {Spahr}}]{Larson1998}
{Larson}, S., {Brownlee}, J., {Hergenrother}, C., \& {Spahr}, T. 1998, in
  Bulletin of the American Astronomical Society, Vol.~30, Bulletin of the
  American Astronomical Society, 1037

\bibitem[{{Lin} {et~al.}(2015){Lin}, {Chen}, {Holman}, \&
  {Ip}}]{2015DPS....4721117L}
{Lin}, H.-W., {Chen}, Y.~T., {Holman}, M.~J., \& {Ip}, W.-H. 2015, in
  AAS/Division for Planetary Sciences Meeting Abstracts, Vol.~47, AAS/Division
  for Planetary Sciences Meeting Abstracts, 211.17

\bibitem[{{Magnier}(2006)}]{Magnier2006}
{Magnier}, E. 2006, in The Advanced Maui Optical and Space Surveillance
  Technologies Conference

\bibitem[{{Morbidelli} {et~al.}(2009){Morbidelli}, {Bottke}, {Nesvorn{\'y}}, \&
  {Levison}}]{Morbidelli2009}
{Morbidelli}, A., {Bottke}, W.~F., {Nesvorn{\'y}}, D., \& {Levison}, H.~F.
  2009, \icarus, 204, 558

\bibitem[{{Petit} {et~al.}(2006){Petit}, {Holman}, {Gladman}, {Kavelaars},
  {Scholl}, \& {Loredo}}]{Petit2006}
{Petit}, J.-M., {Holman}, M.~J., {Gladman}, B.~J., {et~al.} 2006, \mnras, 365,
  429

\bibitem[{{Petit} {et~al.}(2008){Petit}, {Kavelaars}, {Gladman}, \&
  {Loredo}}]{Petit2008}
{Petit}, J.-M., {Kavelaars}, J.~J., {Gladman}, B., \& {Loredo}, T. 2008, {Size
  Distribution of Multikilometer Transneptunian Objects}, ed. M.~A. {Barucci},
  H.~{Boehnhardt}, D.~P. {Cruikshank}, A.~{Morbidelli}, \& R.~{Dotson}, 71--87

\bibitem[{{Schlichting} {et~al.}(2013){Schlichting}, {Fuentes}, \&
  {Trilling}}]{2013AJ....146...36S}
{Schlichting}, H.~E., {Fuentes}, C.~I., \& {Trilling}, D.~E. 2013, \aj, 146, 36

\bibitem[{{Schlichting} {et~al.}(2012){Schlichting}, {Ofek}, {Sari}, {Nelan},
  {Gal-Yam}, {Wenz}, {Muirhead}, {Javanfar}, \& {Livio}}]{Schlichting2012}
{Schlichting}, H.~E., {Ofek}, E.~O., {Sari}, R., {et~al.} 2012, \apj, 761, 150

\bibitem[{{Shankman} {et~al.}(2013){Shankman}, {Gladman}, {Kaib}, {Kavelaars},
  \& {Petit}}]{Shankman2013}
{Shankman}, C., {Gladman}, B.~J., {Kaib}, N., {Kavelaars}, J.~J., \& {Petit},
  J.~M. 2013, \apjl, 764, L2

\bibitem[{{Sheppard} \& {Trujillo}(2010)}]{Sheppard2010b}
{Sheppard}, S.~S., \& {Trujillo}, C.~A. 2010, \apjl, 723, L233

\bibitem[{{Sheppard} {et~al.}(2011){Sheppard}, {Udalski}, {Trujillo}, {Kubiak},
  {Pietrzynski}, {Poleski}, {Soszynski}, {Szyma{\'n}ski}, \&
  {Ulaczyk}}]{Sheppard2011}
{Sheppard}, S.~S., {Udalski}, A., {Trujillo}, C., {et~al.} 2011, \aj, 142, 98

\bibitem[{{Trujillo} \& {Sheppard}(2014)}]{Trujillo2014}
{Trujillo}, C.~A., \& {Sheppard}, S.~S. 2014, \nat, 507, 471

\bibitem[{{Wainscoat} {et~al.}(2015)}]{Wainscoat2015}
{Wainscoat}, R., {et~al.} 2015, in IAU Symposium 318, ed. S.~Chesley \&
  R.~Jedicke, 293

\end{thebibliography}

\end{document}